\documentclass[twocolumn,floats,showpacs,prb]{revtex4}

\usepackage{graphicx}
\usepackage{amsmath}
\usepackage{amsfonts}
\usepackage{amsbsy}
\usepackage{bbm}

\begin{document}

\title{Spin Hall effect and Zitterbewegung in an electron waveguide}

\author{P. Brusheim}
\email{Patrik.Brusheim@ftf.lth.se}
\author{H. Q. Xu}
\thanks{Corresponding author} \email{Hongqi.Xu@ftf.lth.se}
\affiliation{ Division of Solid State Physics, Lund University, Box 118, S-22100 Lund, Sweden.}

\date{\today}
\begin{abstract}
We study spin-resolved probability distributions for electrons in a multi-channel waveguide in the presence of spin-orbit interaction. For a spin-polarized electron injection, a Zitterbewegung pattern is predicted in the probability distribution of the electrons in the waveguide. For a spin-unpolarized injection, the spin-resolved electron probability in the waveguide shows spin accumulations.  In addition to the spin Hall phenomenon, namely accumulations of opposite spins at the lateral edges of the waveguide, we predict the existence of a regular stripe pattern of spin accumulations in the internal region of the waveguide. We show that the predicted Zitterbewegung and spin Hall effect stem from the same mechanism and are formed from coherent states of electrons in the waveguide.
\end{abstract}


\maketitle

Spin-orbit interaction (SOI) is a relativistic effect: it arises from the fact that an electron moving with a finite momentum in a non-uniform electric
potential will see an effective magnetic field acting on its spin. The SOI plays an important role in the field of spintronics since it is in most cases responsible for 
spin rotation and relaxation. Recently an increased
interest has been focused on a consequence of the SOI, namely a spatial spin accumulation in a two-dimensional system in the absence of a magnetic field, known as the spin Hall effect.\cite{Hirsch99,SZhang00,Sheng05,Hankiewicz04,Dyakonov71,Kato04,Sih05,Wunderlich05,
  Nikolic05,Nikolic05f,Sinova04,Mishchenko04,Rashba04,Inoue04,Chalaev05,Khaetskii06,Reynoso06_05} Another SOI-induced phenomenon of current interest is the
\emph{Zitterbewegung} of electrons in semiconductors,\cite{Schliemann05,Lee05,Jiang05} in analogy
with the oscillating behavior of a free relativistic electron due to interference
between its positive- and negative-energy state components. The previous theoretical studies were focused 
on two-dimensional systems or quasi-one-dimensional systems with electron
injection in the lowest subband only. However, in a realistic experimental
setup the Fermi energy and width of the sample are often set at values for which the electron transport is in the multi-subband (multi-channel) regime. It is known that the SOI will induce interaction between subbands as well as the spin states.\cite{Mireles01,Wang04,Zhang05} These interactions will bring an
initially prepared spin state into a quantum-coherent spin-mixed state and
can give rise to complicated spin-dependent electron transport phenomena. To
capture all the physics of the problem an exact multi-subband treatment is
hence necessary. 

It has been theoretically shown that the vertex corrections will cancel the
transverse spin current in an infinite
system.\cite{Mishchenko04,Rashba04,Inoue04,Chalaev05,Khaetskii06} However, the
disappearance of a transverse spin current does not exclude the existence of
spatial spin density modulations arising as a coherent interference effect. Reynoso, Usaj and Balseiro have recently studied the spin Hall effect as a
coherent phenomena in a finite system by calculating the expectation value of the spin operators.\cite{Reynoso06_05} In this paper we
will calculate the wavefunction and its individual spin components to study the intrinsic spin-Hall effect
and Zitterbewegung in a multi-channel electron waveguide in the absence of an external magnetic field. We will show that the
Zitterbewegung and the intrinsic spin Hall effect is essentially the same kind of
phenomena. We will further show the existence of a reversal of the
direction of spatially accumulated spins for electrons injected from individual channels in the waveguide, which can not be
explained by a simple heuristic force operator consideration previously adopted
in the literature. We will also show that regular stripe patterns of spin accumulations are formed inside the multi-channel waveguide in
addition to the usual edge spin accumulations found in the spin Hall
effect. These patterns are shown to arise from coherent states in the waveguide and the number of stripes in the patterns are found to be related to the number of open channels in the leads.

To study the problem we consider a two-dimensional electron gas (2DEG) formed in the $x$-$y$ plane of a semiconductor heterostructure. The 2DEG is restricted to a waveguide of width $w$ by a hard-wall confinement potential $U_c(y)$, i.e., $U_c(y)=0$ for $y\in[-w/2,w/2]$ and $\infty$ otherwise, in the transverse $y$-direction. The structure is subjected to an electrical field along the unit vector $\boldsymbol{\xi}$ giving rise to a Rashba SOI of strength $\alpha$. For crystal structures without inversion symmetry, there will in addition be a Dresselhaus SOI contribution, characterized by the strength $\beta$. Such inversion asymmetry is found in, e.g., zinc-blende crystals of III-V materials. The waveguide is connected with two perfect leads with vanishing SOI. The electrons are injected from one of the two leads into the region with a finite SOI and transport coherently in the longitudinal $x$-direction. The model Hamiltonian describing such a system under the effective mass approximation has the form,
\begin{eqnarray}
  H &=& \boldsymbol{\sigma}_0\left[\frac{\mathbf{p}^2}{2m^*} + U_c(y)\right] + \frac{1}{2\hbar}\Big[\alpha(\mathbf{r})\boldsymbol{\sigma}\cdot \left(\mathbf{p}\times\boldsymbol{\xi}\right) \nonumber \\
    &+& \beta(\mathbf{r})\left(\sigma_x p_x - \sigma_y p_y\right) + H.c.\Big],
\end{eqnarray}
where $\mathbf{p}$ is the momentum operator, $m^*$ the electron effective mass, which is
taken to be $m^*=0.042\, m_e$ corresponding to an InGaAs quantum-well system,
$\boldsymbol{\sigma}=(\sigma_x, \sigma_y, \sigma_z)$ the vector of the Pauli
matrices, and $\boldsymbol{\sigma}_0$ the unit matrix. The inclusion of the SOI breaks the reflection invariance of the system in the transverse direction, i.e., $[R_y,H]\neq0$. However, for a symmetric confining potential the system is invariant under operation of $\sigma_y R_y$, i.e., $[\sigma_y R_y, H]=0$. This symmetry has an important implication: for a spin-unpolarized electron injection, the spin-up electron probability distribution in the system is exactly the mirror image of the spin-down electron probability distribution with respect to the transverse reflection $R_y$. Thus, if an accumulation of spin occurs on one side of the waveguide, the same amount of accumulation of the opposite spin will occur on the other side of the waveguide. Since an unpolarized electron injection beam is an
incoherent mixture of two spin injections, this symmetry asserts that all information about the
probability distribution, regardless of the injection condition, is contained
in the probability distribution of a polarized injection beam. One needs only to
consider the calculations for one spin injection, since the results from the other spin injection follows from the symmetry requirement.

The (spin-unpolarized) probability distribution, $\rho^{\gamma}(x,y)$, for electrons injected from a lead in a spin-$\gamma$ state can be obtained from 
\begin{equation}
  \rho^{\gamma}(x,y) =\sum_{q\sigma}\rho_{q}^{\sigma\gamma}(x,y) \sim \sum_{\sigma}\sum_{\{q | k_q\in\mathbb{R}\}} \frac{|\langle \chi(\sigma)|\Psi_q^{\gamma}(x,y)\rangle|^2}{k_q},
\end{equation}
where $\Psi_q^{\gamma}(x,y)$ is the wave function of an electron injected from the lead in the $q$th subband with spin $\gamma\!=\,\uparrow$ or $\downarrow$, $|\chi(\gamma\!=\,\uparrow)\rangle=(1,0)^T$ and $|\chi(\gamma\!=\,\downarrow)\rangle=(0,1)^T$ are the two spin states with the spin quantization axis along the electric field direction, $k_q=\left[2m*(E_F-E_q)/\hbar^2\right]^{1/2}$ with $E_F$ being the Fermi
energy and $E_q$ the $q$th subband energy in the injection lead, and
$\rho^{\sigma\gamma}(x,y)$ is the spin-$\sigma$ probability
distribution. In
the present work, the wave function $\Psi_q^{\gamma}$ has been calculated by
employing an exact, spin-dependent, multi-mode scattering matrix technique (see Refs.~\onlinecite{Zhang05,Xu9495,Ko88} for the detail of the theoretical method). 

\begin{figure}[tb]
  \begin{center}
    \includegraphics[scale=0.28]{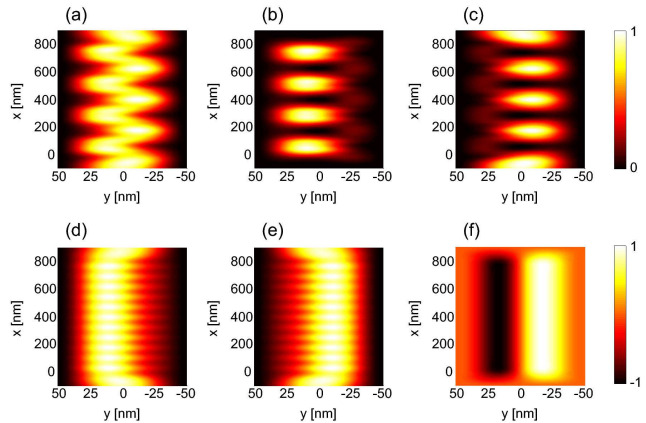}
    \caption{(Color online) Normalized probability distribution for electrons injected in a spin-up state (a)-(b) and for a spin-unpolarized electron injection (d)-(f). (a) is for probability distribution, $\rho_1^{\uparrow}$; (b) for its spin-down projection, $\rho_1^{\downarrow\uparrow}$; (c) for its spin-up projection, $\rho_1^{\uparrow\uparrow}$; (d) for the probability distribution of spin-down electrons in the spin-unpolarized injection, $\rho_1^{\downarrow\uparrow} + \rho_1^{\downarrow\downarrow}$; (e) for the probability distribution of spin-up electrons in the spin-unpolarized injection, $\rho_1^{\uparrow\uparrow} + \rho_1^{\uparrow\downarrow}$; and (f) for the spin polarization distribution in the spin-unpolarized electron injection, $\rho_{_1,pol}=\rho_1^{\uparrow\uparrow} + \rho_1^{\uparrow\downarrow} - \rho_1^{\downarrow\uparrow} - \rho_1^{\downarrow\downarrow}$. In all the figures, the channel width is set at $w=100$ nm and the Rashba SOI strength is $\alpha=3\times10^{-11}$ eVm for $x\in[0,800]$ nm and is decreased adiabatically down to zero in $x\in[-100,0] \cup [800,900]$ nm. The Fermi energy is set at $E_F =2$ meV. The upper scale bar shows the measure for plots (a)--(e), while the lower scale bar shows the measure for plot (f).}
    \label{Fig1}
  \end{center}
\end{figure}

Consider now the case when only the Rashba SOI is present, i.e., $\beta=0$ and
$\alpha\neq 0$, in Eq.~(1). Since $[\sigma_i,H]\neq 0$ for all $i$, an electron initially
prepared in a pure spin state in the lead will evolve into a quantum-coherently spin-mixed  
state as it travels in the waveguide with a finite SOI. This coherent
spin evolution will give rise to a spin density modulation along the waveguide. To visualize these properties, we calculate the electron probability distributions in the waveguide. We consider the common situation with the electric field, $\boldsymbol{\xi}$, set along the $z$-direction. In this case, the Rashba SOI term can be written as $(\alpha/2\hbar)(\sigma_x p_y-\sigma_y p_x) + H.c.$ 

We first assume the width of the waveguide $w=100$ nm and the Fermi energy
$E_F=2$ meV, corresponding to the case where only one channel is open in the
leads. The SOI strength is taken to be $\alpha=3\times10^{-11}$ eVm in the
region with the full strength of the Rashba SOI, defined in $x\in[0,800]$ nm,
and decreased adiabatically down to zero in the transition regions of the
entrance and exit, defined in $x\in[-100,0]$ and $x\in[800,900]$ nm, respectively. Figure~\ref{Fig1}(a) shows the calculated total (charge) probability distribution
$\rho^{\uparrow}=\rho^{\uparrow\uparrow} + \rho^{\downarrow\uparrow}$ for
electrons injected in a pure spin-up state from the lower lead. It is seen that the electron wave function exhibits a transversely oscillating behavior along the
waveguide. This is the Zitterbewegung arising from interference between the two
spin components of the electron state. By considering the spin-up [Fig~\ref{Fig1}(b)]
and the spin-down [Fig~\ref{Fig1}(c)] projections separately, spin-wave
patterns, in the form of spatially separated localized islands of
high spin distribution density, are found. The two probability distribution
components, $\rho^{\uparrow\downarrow}$ and $\rho^{\downarrow\downarrow}$,
for the spin-down injection can be obtained from mirror reflection of the results shown in
Figs.~\ref{Fig1}(a)--\ref{Fig1}(c) as required by $[\sigma_yR_y,H]=0$. It is evident that the spin-up and spin-down electron wave-function components are localized in different
sides of the waveguide. This accumulation of different spins along opposite
edges of the waveguide is a fundamental property of the spin Hall effect observed
recently. The spin projected electron probability distributions for a spin-unpolarized
injection are shown in Figs.~\ref{Fig1}(d) and \ref{Fig1}(e). Due to the symmetry
of the system under the operation of $\sigma_y R_y$, the total distribution will be symmetrical in the transverse direction, and the spin-up (spin-down) probability distribution, Fig.~\ref{Fig1}(d) [Fig.~\ref{Fig1}(e)], is the mirror image of the corresponding spin-down (spin-up) probability distribution. Hence no spin-polarization will be detected in the measured two-terminal conductance. This was rigorously shown, in Refs.~\onlinecite{Zhai05} and
\onlinecite{Bulgakov02}, to be always hold, independent of the details of the
conductor, when only one channel is open for conduction in the leads. Accumulations of electron spins with opposite polarizations at the opposite edges of a wide conductor have recently
been experimentally observed \cite{Kato04,Sih05,Wunderlich05}. What is measured
in such experiments is the net spin density distribution, $\rho_{pol}=\rho^{\uparrow\uparrow}+\rho^{\uparrow\downarrow}-\rho^{\downarrow\uparrow}-\rho^{\downarrow\downarrow}$, in the sample for a spin-unpolarized electron injection. Figure~\ref{Fig1}(f) displays the results of calculations for $\rho_{pol}$ in the waveguide and shows clearly the signature of the spin Hall effect. It is seen that the spin Hall effect and the Zitterbewegung is essentially the same kind of phenomena, since one follows from the other, depending on if only one or both spin states are considered in the injection source and whether the spin accumulation or the charge accumulation is measured. In other words, the results of Figs.~\ref{Fig1}(a), \ref{Fig1}(d), \ref{Fig1}(e) and \ref{Fig1}(f) can be constructed from
Figs.~\ref{Fig1}(b) and \ref{Fig1}(c) and their mirror images. 

\begin{figure}[tb]
  \begin{center}
    \includegraphics[scale=0.28]{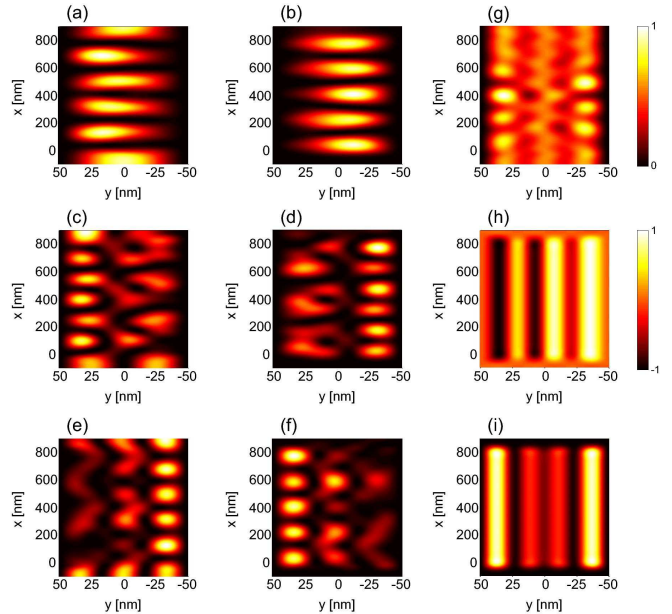}
    \caption{(Color online) Normalized electron probability distribution and spin-projected electron probability distribution for spin-polarized and spin-unpolarized injections in the same waveguide as in Fig.~\ref{Fig1} at the Fermi energy $E_F =12$ meV (the case with three opening channels in the leads). (a) is for $\rho_1^{\uparrow\uparrow}$; (b) for
    $\rho_1^{\downarrow\uparrow}$; (c) for $\rho_2^{\uparrow\uparrow}$; (d) for
    $\rho_2^{\downarrow\uparrow}$; (e) for $\rho_3^{\uparrow\uparrow}$; (f) for
    $\rho_3^{\downarrow\uparrow}$; (g) is the the normalized total probability
    distribution for spin-up electron injection, $\rho^{\uparrow}= \sum_{q=1}^{3}(\rho_q^{\uparrow\uparrow} + \rho_q^{\downarrow\uparrow})$; (h) is the normalized total 
    spin distribution for spin-unpolarized electron injection,
    $\rho_{pol}=\sum_{q=1}^{3}(\rho_q^{\uparrow\uparrow} + \rho_q^{\uparrow\downarrow} - \rho_q^{\downarrow\uparrow} - \rho_q^{\downarrow\downarrow})$; and (i) the same as (h) but for the distribution of the spin polarization along the transverse $y$ direction.  The upper scale bar shows the measure for plots (a)--(g) and (i), while the lower scale bar shows the measure for plot (h).}
    \label{Fig2}
  \end{center}
\end{figure}

When the Fermi energy is increased, more channels are open for
conduction in the leads. Figure~\ref{Fig2} shows the probability distributions
for the Fermi energy set at $E_F=12$ meV, for which three spin-degenerate channels are open for conduction in the leads. In Figs.~\ref{Fig2}(a)-\ref{Fig2}(f), the spin probability distributions for spin-up electrons injected from the lower lead in the three individual channels are
shown. The probability distributions for the spin-up injection in the lowest channel,
$\rho_1^{\uparrow\uparrow}$ and $\rho_1^{\downarrow\uparrow}$, shown in
Figs.~\ref{Fig2}(a) and \ref{Fig2}(b) exhibit similar Zitterbewegung and spin-Hall
patterns as was found in the single-channel system. However, here a reversal of the spin-polarization direction is found:
the spin-up (spin-down) probability density is localized along the
left (right) edge of the waveguide. This is also true for the probability
distributions for the spin-up injection in the second lowest channel,
$\rho_2^{\uparrow\uparrow}$ and $\rho_2^{\downarrow\uparrow}$, as shown in
Figs.~\ref{Fig2}(c) and \ref{Fig2}(d). Only the probability distributions for the spin-up injection in the highest subband, $\rho_3^{\uparrow\uparrow}$ and
$\rho_3^{\downarrow\uparrow}$, shown in Figs.~\ref{Fig2}(e) and (f), exhibit the same
spin parity as in the single-channel opening case. We note that this phenomenon can not be
explained by the force operator, derived from the Hamiltonian, as in, e.g., Ref~\onlinecite{Nikolic05f}, and implies that such a heuristic approach
fails to explain the full quantum mechanical calculations. Indeed,
for equilibrium systems considered here, no physical force is present and
the observed Zitterbewegung and spin-Hall effect are non-dynamical but merely coherent phenomena. 

In light of these findings, one may ask if the
Zitterbewegung and spin-Hall effect can still be observed in the multi-channel
system when the contributions from all the individual channel injections are taken into account. To answer this question we show the total probability distribution,
$\rho^{\uparrow}$, for a spin-up electron injection in the waveguide in  Fig.~\ref{Fig2}(g) and the spin polarization distribution, $\rho_{pol}$, for a spin-unpolarized electron injection in Fig.~\ref{Fig2}(h). It is seen in Fig.~\ref{Fig2}(g) that Zitterbewegung oscillations are still found along
  the waveguide, albeit more complicated as compared to the single-channel opening
  case. The spin polarization distribution, $\rho_{pol}$, shown in
  Fig.~\ref{Fig2}(h) exhibits a regular spin density pattern. Strong spin
  polarization with opposite sign is found at the two edges of the waveguide,
  as observed in recent experiments. However, spin accumulation, with a
  pattern of alternative spin-polarization stripes, also occurs inside the
  waveguide. The number of stripes (including the edge spin polarization stripe) for each spin-polarization direction is equal to the number of the open channels in the waveguide. Experimental observation of the interesting internal spin-polarization structure is certainly challenging; it requires a spin detection setup with a high spatial resolution (about 50 nm or better for the system discussed in this work). However, it should be noted that no spin Hall effect could be observed if the spin polarization along a direction perpendicular to the electric field $z$ direction is measured in the system. Although the spin polarization along the transverse $y$ direction shows a regular stripe pattern of accumulation, the same spin polarity is seen in all the stripes, as is shown in Fig.~\ref{Fig2}(i). This can be understood as a result of polarization (or magnetization) by the effective transverse magnetic field arising from the Rashba term. 

\begin{figure}[tb]
  \begin{center}
    \includegraphics[scale=0.40]{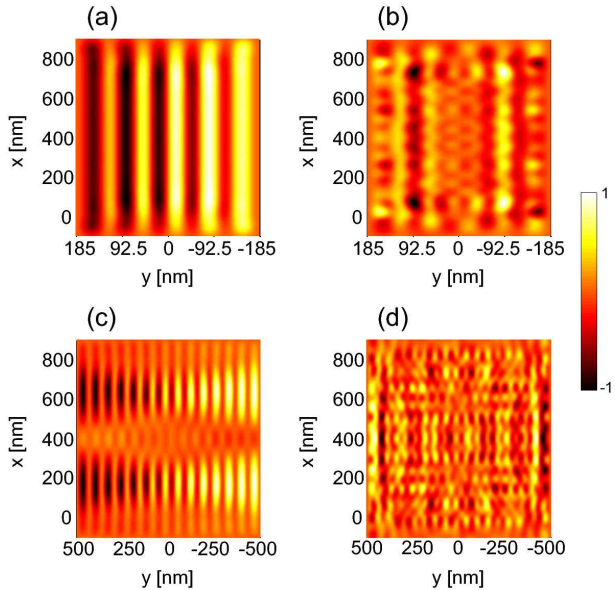}
    \caption{(Color online) Normalized spin probability distribution, $\rho_{pol}$,
    for spin-unpolarized electron injection at the Fermi energy $E_F=2$ meV in a waveguide with (a) $w=370$ nm and
    $\alpha=1\times10^{-11}$ eVm, (b) $w=370$ nm and $\alpha=3\times10^{-11}$ eVm,
    (c) $w=1000$ nm and $\alpha=0.1\times10^{-11}$ eVm, and (d) $w=1000$ nm and
    $\alpha=3\times10^{-11}$. The other waveguide parameters assumed are the same as in Figs.~\ref{Fig1} and \ref{Fig2}.}
    \label{Fig3}
  \end{center}
\end{figure}

For a wide conductor, a large number of channels are typically open for
conduction and the inter-subband mixing induced by the SOI can produce a
complicated pattern in the electron probability distribution. In Fig.~\ref{Fig3} we
show the calculated spin polarization $\rho_{pol}$ for a spin-unpolarized electron injection at the Fermi energy $E_F=2$ meV in a waveguide of width $w=370$ nm (with five open channels) and a waveguide of width $w=1$ $\mu$m (with 14 open channels) at weak and strong SOI
strengths. The distributions are symmetric with respect to $\sigma_yR_y$. For
the weak SOI strength, the same structural patterns of alternative spin-polarization stripes [Figs~\ref{Fig3}(a) and \ref{Fig3}(c)]
as observed in the narrower waveguide are found. For the strong SOI strength, the spin-polarization distribution patterns [Figs.~\ref{Fig3}(b) and \ref{Fig3}(d)] show complex structures. Although spatial spin accumulations are observable even under these circumstances, regular spin accumulations at the edges of the waveguides and internal stripe structures are destroyed by the strong SOI induced inter-channel scattering. This result indicates that the regular spin Hall effect can only be clearly observed in a range of SOI strengths in a wide, multi-channel electron waveguide.

Since the Rashba and the Dresselhaus term in the Hamiltonian are related
through a unitary transformation, with the substitutions of $\sigma_x \rightarrow \sigma_y$, $\sigma_y \rightarrow \sigma_x$ and $z \rightarrow -z$, the fundamental results presented
above hold for the case of $\alpha=0$ and $\beta\neq 0$, i.e., when only the Dresselhaus SOI is present. For $\alpha=\pm\beta$ the spin-dependent part of the Hamiltonian takes the form  
\begin{equation}
  H_{\pm}^s = \frac{\alpha}{\hbar}\left(p_y \pm p_x\right)\boldsymbol{\sigma}\cdot\left(\mathbf{e}_x \mp \mathbf{e}_y\right).
\end{equation}
The SOI induced effective magnetic field direction is in this case independent of the
momentum. This means that a well-defined spin quantization axis can be found
throughout the SOI region and spin is therefore a conserved quantity. In Ref.~\onlinecite{Miller03}, it
was shown that in III-V systems the Rashba and Dresselhaus SOIs can be of
comparable strengths. This
would imply that no Zitterbewegung~\cite{Schliemann05} nor spin-Hall
effect could occur in this situation.   

In conclusion, we have studied the total and spin-resolved electron probability distributions in a ballistic waveguide with SOI. For a spin-polarized electron injection, a Zitterbewegung pattern is observed in the probability distribution of the electrons in the waveguide. For a spin-unpolarized injection, the spin-resolved electron probability in the waveguide shows spin accumulations.  The intrinsic spin Hall effect, namely accumulations of opposite spins at the lateral edges of the waveguide, can be observed in the SOI region.  We have also predicted the existence of a regular stripe pattern of spin accumulations inside the waveguide. Finally we have shown that the Zitterbewegung and the spin Hall effect found in this work stem from the same mechanism and are formed from the coherent states of electrons in the waveguide.

This work was supported by the Swedish Research Council (VR) and by the Swedish Foundation for Strategic Research (SSF) through the Nanometer Structure Consortium at Lund University.

\end{document}